\documentclass[reprint,superscriptaddress,showkeys,amsmath,amssymb,prl,longbibliography,citeautoscript,aps]{revtex4-2}

\usepackage[dvipsnames]{xcolor}

\usepackage{amsmath}
\usepackage{amssymb}
\usepackage[hidelinks]{hyperref}
\usepackage{chemformula}
\usepackage{siunitx}
\usepackage[capitalise]{cleveref}
\usepackage{soul}

\makeatletter
\def\maketitle{
\@author@finish
\title@column\titleblock@produce
\suppressfloats[t]}
\makeatother

\begin{document}

\title{Revealing Fast Ionic Conduction in Solid Electrolytes through Machine Learning Accelerated Raman Calculations}

\author{Manuel Grumet}
\affiliation{Physics Department, TUM School of Natural Sciences, Technical University of Munich, 85748 Garching, Germany}
\affiliation{Atomistic Modeling Center, Munich Data Science Institute, Technical University of Munich, Germany}
\author{Takeru Miyagawa}
\affiliation{Physics Department, TUM School of Natural Sciences, Technical University of Munich, 85748 Garching, Germany}
\author{Olivier Pittet}
\affiliation{Physics Department, TUM School of Natural Sciences, Technical University of Munich, 85748 Garching, Germany}
\affiliation{Laboratory of Computational Science and Modeling, Institut des Matériaux,
École Polytechnique Fédérale de Lausanne, 1015 Lausanne, Switzerland}
\author{Paolo Pegolo}
\affiliation{Laboratory of Computational Science and Modeling, Institut des Matériaux,
École Polytechnique Fédérale de Lausanne, 1015 Lausanne, Switzerland}
\author{Karin S. Thalmann}
\altaffiliation{Present address: Institute of Physics, University of Freiburg, 79104 Freiburg, Germany}
\affiliation{Physics Department, TUM School of Natural Sciences, Technical University of Munich, 85748 Garching, Germany}
\author{Waldemar Kaiser}
\affiliation{Physics Department, TUM School of Natural Sciences, Technical University of Munich, 85748 Garching, Germany}
\author{David A. Egger}
\email{david.egger@tum.de}
\affiliation{Physics Department, TUM School of Natural Sciences, Technical University of Munich, 85748 Garching, Germany}
\affiliation{Atomistic Modeling Center, Munich Data Science Institute, Technical University of Munich, Germany}

\begin{abstract}
Fast ionic conduction is a defining property of solid electrolytes for all-solid-state batteries.
Previous studies have suggested that liquid-like cation motion associated with fast ionic transport can disrupt crystalline symmetry, thereby lifting Raman selection rules.
Here, we exploit the resulting low-frequency, diffusive Raman scattering as a spectral signature of fast ionic conduction and develop a machine learning-accelerated computational pipeline to identify promising solid electrolytes based on this feature.
By overcoming the steep computational barriers to calculating Raman spectra of strongly disordered materials at finite temperatures, we achieve near-ab initio accuracy and demonstrate the predictive power of our approach for sodium-ion conductors, revealing clear Raman signatures of liquid-like ion conduction.
This work highlights how machine learning can bridge atomistic simulations and experimental observables, enabling data-efficient discovery of fast-ion conductors.

\end{abstract}

\maketitle

\noindent


The development of sustainable energy technologies relies on the discovery and optimization of functional materials for efficient energy storage and conversion.
Solid electrolytes play a central role in this effort as they enable all-solid-state batteries with high energy density and thermal stability~\cite{zhang_etal_2018, famprikis_etal_2019, janek_zeier_2023}. 
Fast conduction of ions through a crystalline host lattice is a key property of solid electrolytes~\cite{he2017origin, ohno2020materials, poletayev2022defect, jun2024diffusion}. 
However, identifying compounds with high ionic conductivity across the vast chemical phase space remains challenging, primarily because synthesis, optimization, and characterization of these materials is time-intensive~\cite{balaish2021processing, hei2024electrolyte}. 
Accelerating the discovery of fast ion-conducting solid electrolytes requires a concerted integration of synthesis control, rapid characterization techniques, and high-throughput computational screening~\cite{jain2016computational, butler2018machine, zhu2021lithium, janek_zeier_2023}. 
Yet, various characterization strategies currently rely on fully assembled device stacks incorporating electrodes and electrolyte/electrode interfaces~\cite{janek2016solid, vadhva2021electrochemical}, which can obscure intrinsic ion transport properties. 
This is especially because the formation of space charge layers at interfaces can modify the ionic conductivity~\cite{dudney1985effect, maier1995ionic}, for instance by introducing additional degradation pathways under applied bias~\cite{richards2016interface, xiao2020understanding}. 

In search for less invasive and more rapid characterization strategies, Raman spectroscopy emerged as a potentially promising tool for identifying fast-ion conductors through characteristic vibrational signatures~\cite{krauskopf2018comparing, brenner_etal_2020, famprikis_etal_2021,  brenner_etal_2022}.
In particular, the emergence of a low-frequency diffusive Raman feature, often referred to as the Raman central peak, has been proposed as a hallmark of fast ionic conduction in a study including one of us~\cite{brenner_etal_2020}.
Work by Delaire and co-workers support this interpretation, showing that in superionic conductors such as \ch{AgCrSe2}, \ch{Li6PS5Cl}, and \ch{Na3PS4}, the onset of fast-ion conduction is accompanied by pronounced soft-mode activity and quasielastic neutron-scattering features analogous to the Raman central peak~\cite{ding_etal_2020, gupta_etal_2021, ding2025liquid}.
If this hypothesis holds, then a Raman central peak could serve as a predictor enabling rapid, non-invasive, and scalable screening of fast-ion conductors.
Such an approach is also appealing from a first-principles computational standpoint, where Raman spectra including central peaks can be predicted from combining molecular dynamics (MD) with time-dependent simulation of dielectric material behavior~\cite{thomas_etal_2013, yaffe2017local, gao2021metal, caicedo-davila_etal_2024}.
Although these predictions are computationally intensive because evaluating the polarizability tensor, $\boldsymbol{\alpha}$, requires repeated density functional perturbation theory (DFPT) calculations, machine learning (ML) approaches now offer efficient surrogates that dramatically reduce this cost~\cite{egger2025machine}.

\begin{figure*}[ht]
    \centering
    \includegraphics[width=0.6\textwidth]{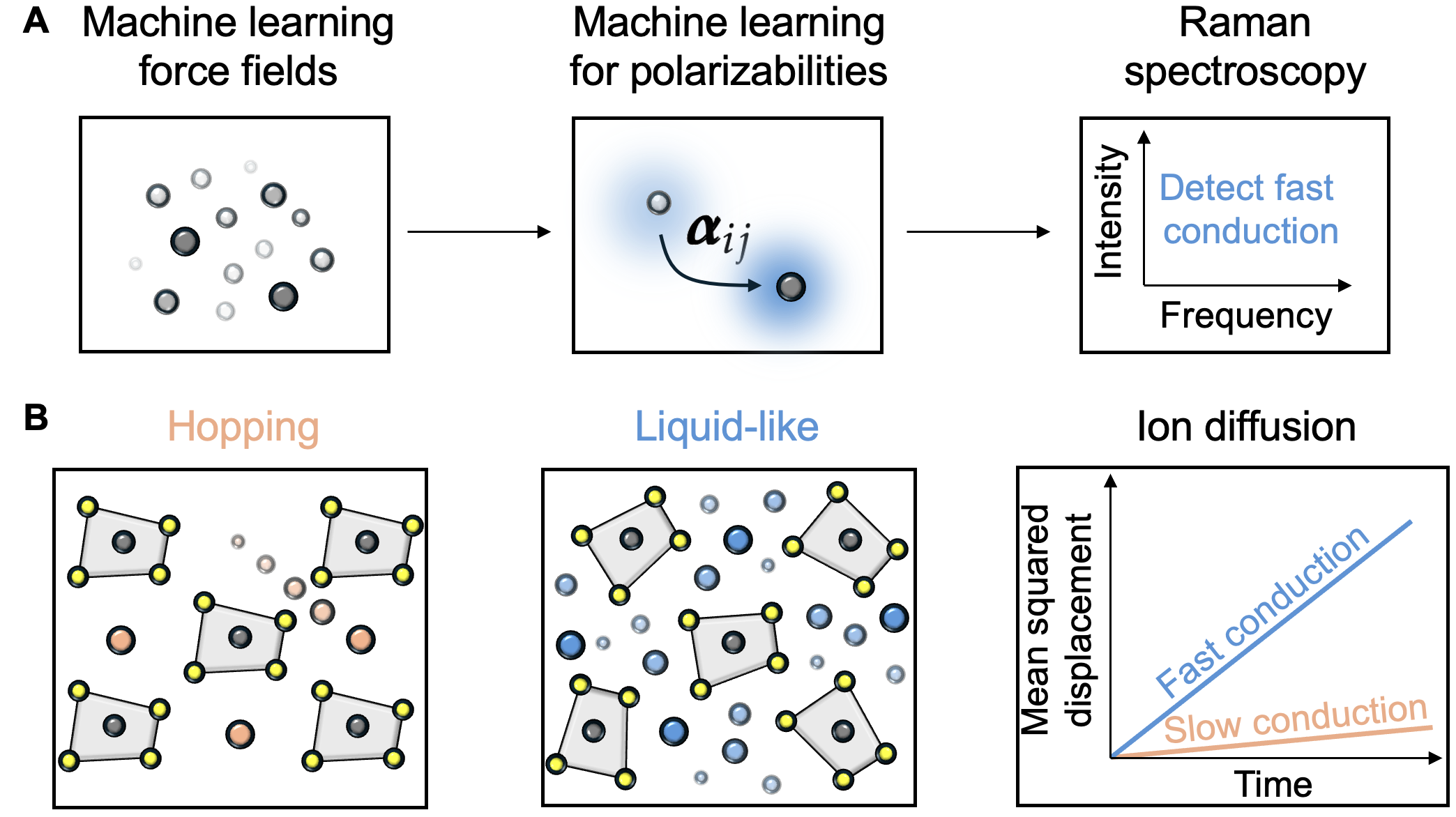}
    \caption{
      \textbf{Machine learning-enabled insights into fast-ion conduction.} (A) Workflow integrating machine-learned force fields and polarizabilities to simulate vibrational spectra and identify signatures of fast-ion conduction using Raman spectroscopy. 
      (B) Schematic comparison of two ion transport regimes: thermally activated hopping (left) versus liquid-like fast conduction (middle), characterized by distinct structural dynamics and diffusion behavior. 
      Mean squared displacement (right) distinguishes fast and slow conduction regimes based on their time dependence.
    }
    \label{fig:1}
\end{figure*}

Building on this perspective, we here propose an efficient computational pipeline to reveal fast ionic conduction in solid electrolytes through Raman spectroscopy.
Specifically, we develop a workflow that combines ML force fields (MLFFs) with ML models for predicting dynamic changes of $\boldsymbol{\alpha}$ to enable efficient computations of Raman signatures (see Fig.~\ref{fig:1}A).
We investigate whether our workflow is capable of detecting symmetry-breaking mechanisms that lead to low-frequency diffusive Raman scattering serving as a predictor to distinguish fast, liquid-like ionic conduction from slower, hopping-like conduction (see Fig.~\ref{fig:1}B).
To this end, we leverage recent advances in ML for tensorial properties and investigate whether it can reach ab-initio accuracy in predicting the Raman spectrum of the prototypical solid electrolyte AgI from DFPT inputs in a data-efficient manner.
Finally, we examine whether the appearance of a Raman central peak appearing in sodium-ion conductors can serve as a signature of liquid-like cation motion and thus enable Raman spectroscopy to identify solid electrolytes with high ionic conductivity.


\begin{figure*}[ht]
    \centering
    \includegraphics[width=0.8\textwidth]{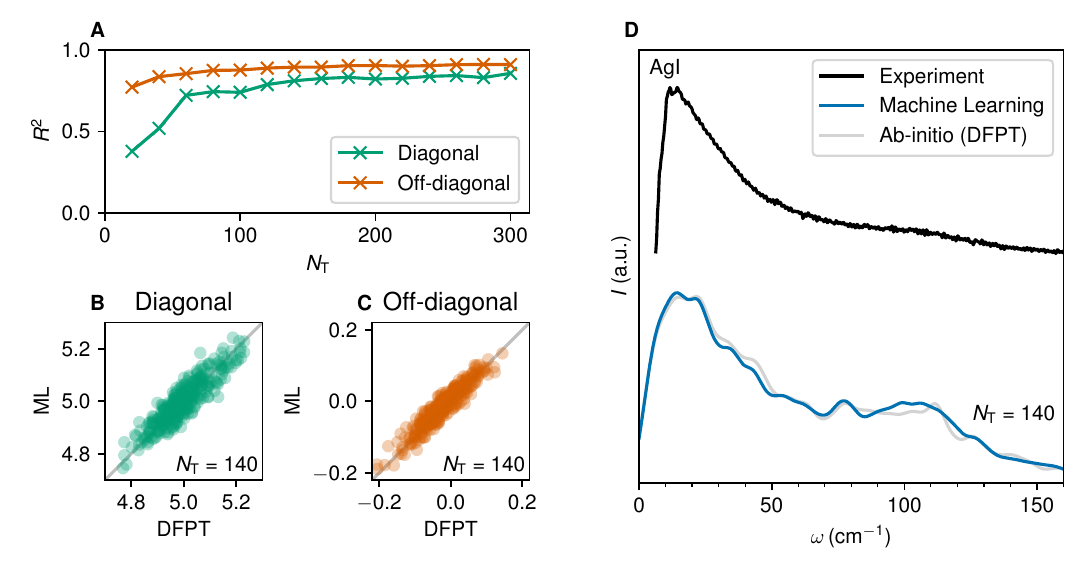}
    \caption{
      \textbf{Performance of the machine-learning approach for AgI.}
      (\textbf{A}) Learning curve showing $R^2$ scores of the machine-learning predictions as a function of the training set size $N_T$.
      (\textbf{B} and \textbf{C}) Scatter plots showing predictions versus \textit{ab initio} values for diagonal and off-diagonal components of the dielectric tensor.
      (\textbf{D}) Computed Raman spectra at 500\,K obtained from both machine-learning-predicted and \textit{ab initio} polarizabilities. An experimentally measured spectrum at 500\,K is shown for comparison \cite{brenner_etal_2020}.
    }
    \label{fig:AgI}
\end{figure*}

We first assess the accuracy of our ML-accelerated Raman simulations for the high-temperature $\alpha$-phase of AgI. 
A prototypical superionic conductor, AgI features a high degree of disorder in its Ag$^+$ sublattice as well as strongly anharmonic host-lattice vibrations \cite{brenner_etal_2020}, which render it a challenging test case for computational modeling.
As described above, our ML-accelerated workflow for Raman computations combines MLFFs with tensorial ML to capture the dynamic evolution of the polarizability tensor, $\boldsymbol{\alpha}$.
We use \texttt{VASP} \cite{kresse_furthmuller_1996a} for generating the MLFFs, as well as for obtaining training data for $\boldsymbol{\alpha}$.
Specifically, the MLFFs are trained ``on-the-fly'' using data generated from density functional theory (DFT) calculations (see Methods section for details). 
In a recent study, some of us demonstrated the high fidelity of such MLFFs for AgI and other solid electrolytes~\cite{miyagawa_etal_2024}.
These previous findings, together with related studies for various solid electrolytes~\cite{huang2021deep, staacke2022tackling, gigli2024mechanism}, lead to the expectation that MLFFs accurately reproduce \textit{ab initio} calculations, which we confirm for the MLFFs used here by comparing MLFF- and DFT-derived forces (see Fig.~\ref{fig:sup_mlff_scatterplots}).

Therefore, while not trivially guaranteed a high accuracy of state-of-the-art MLFFs for complex dynamic materials is expected, once they have been properly trained and tested for the materials under investigation. 
The more challenging and novel aspect of our work lies in learning and predicting $\boldsymbol{\alpha}$ for solid electrolytes. 
Although ML models have achieved high fidelity for predicting $\boldsymbol{\alpha}$ this has not, to our knowledge, been demonstrated for this structurally and dielectrically complex material class.
Accordingly, we learn and predict $\boldsymbol{\alpha}$ using supervised linear regression methods, trained on datasets of polarizability tensors computed via DFPT for structures randomly selected from MLFF runs (see Methods section for details).
Specifically, we employ an equivariant ML approach based on $\lambda$-SOAP descriptors~\cite{grisafi_etal_2018}, implemented using the \texttt{featomic} and \texttt{metatensor} packages~\cite{metatensor}, and investigate their
accuracy in combination with MLFF trajectories first.
Also note that, for the sake of simplicity, we do not distinguish between the polarizability tensor $\boldsymbol{\alpha}$ and the dielectric tensor $\boldsymbol{\epsilon}$ in this work, since both quantities have the same frequency behavior and are therefore interchangeable for the purpose of Raman calculations~\cite{egger2025machine}.

Fig.~\ref{fig:AgI} shows key performance metrics of our ML-accelerated workflow for the case of AgI.
As is the case with many materials, the off-diagonal components of $\boldsymbol{\alpha}$ in AgI are much smaller in magnitude than the diagonal ones.
We emphasize, however, that accurate predictions are important for both because temporal correlations across all components determine the Raman spectrum (see Methods section).
Remarkably, the learning curve (Fig.~\ref{fig:AgI}A) shows that already with relatively modest training set size of $N_\mathrm{T}=140$ samples, the model attains very good accuracy, gauged by $R^2>0.8$ for both diagonal (Fig.~\ref{fig:AgI}B) and off-diagonal (Fig.~\ref{fig:AgI}C) components relative to the DFPT ground truth.
Interestingly, the off-diagonal components show slightly quicker convergence with respect to $N_\mathrm{T}$ and appear reasonably accurate already with using only few tens of samples.
We note in passing that we find improved accuracy in both components 
(see Fig.~\ref{fig:AgI}D) when further increasing $N_\mathrm{T}$, although the Raman spectrum -- our observable of interest -- only improves marginally compared to the ground truth (see Fig.~\ref{fig:sup_training_convergence_AgI}). 

We now turn to the Raman spectrum of AgI at 500\,K, a temperature at which the material has already adopted the ion-conducting $\alpha$-phase.
The experimental Raman spectrum \cite{brenner_etal_2020} reproduced in Fig.~\ref{fig:AgI}D highlights broad and intensive low-frequency features, that is a Raman central peak at around 20\,cm$^{-1}$, followed by a broad shoulder at approximately 90\,cm$^{-1}$.
It is well-known that these Raman features are stemming from the diffusive and strongly anharmonic atomic motions that are present in AgI at this temperature~\cite{boyce1977position, geisel1977liquid,  nemanich1980light,brenner_etal_2020}.
Comparing the experimental spectrum with the computed results in Fig.~\ref{fig:AgI}D, we first note that the DFPT-computed spectrum reproduces the experimental features well.
Importantly, the ML-computed Raman spectrum also shows excellent agreement with the ground-truth spectrum obtained from DFPT (see Fig.~\ref{fig:AgI}D), consistent with our findings above.
Using the challenging and representative test case of AgI, we therefore establish that our ML-accelerated Raman workflow can accurately capture the Raman response arising from strongly diffusive and anharmonic atomic motions in solid electrolytes.

An important question that arises in light of these findings is whether the Raman central peak observed in the $\alpha$-phase of AgI is indicative of fast-ion conduction across multiple material families~\cite{brenner_etal_2020}.
Precedent for this hypothesis stems from the following considerations:
occurrence of a Raman central peak is a signature of anharmonic atomic motions, which dynamically break the crystalline symmetry on the timescale that is relevant to Raman scattering, such that Raman selection rules are relaxed and all vibrations become Raman active.
This leads to close correspondence between the vibrational density of states (VDOS) and the material's Raman response, which we explicitly confirm for $\alpha$-AgI in Fig.~\ref{fig:sup_vdos}.
Because the zero-frequency component of the VDOS reflects long-time ionic dynamics, its magnitude correlates with ionic diffusion such that larger contributions in the VDOS at $\omega=0$ correspond to faster ionic conduction \cite{nitzan_chemical_dynamics_in_condensed_phases}. 
As the zero-frequency VDOS amplitude increases with ionic diffusion, the Raman central peak is expected to exhibit the same correlation once Raman selection rules are relaxed.
A mechanism through which this can occur is liquid-like cation motion in fast-ion conductors, where short residence times of the cations break symmetry locally and dynamically, inducing relaxational motions of host-lattice ions~\cite{brenner_etal_2020}. 
Taken together, these considerations suggest that liquid-like ionic conduction should give rise to an intense low-frequency Raman response.

We note that the same expectation can also be derived by considering the optical conductivity and its zero-frequency component, which reflects the ionic mobility \cite{fulde_etal_1975,sato_etal_1992}.
The optical conductivity is obtained from the current-current correlation function and therefore closely parallels the VDOS, which is derived from the velocity-velocity correlation function.
Thus, when the ionic mobility is high, both the optical conductivity and VDOS exhibit enhanced low-frequency contributions.
In systems where Raman selection rules are lifted due to liquid-like ionic motion, this again translates into pronounced low-frequency Raman intensity.

\begin{figure*}
    \centering
    \includegraphics[width=0.8\textwidth]{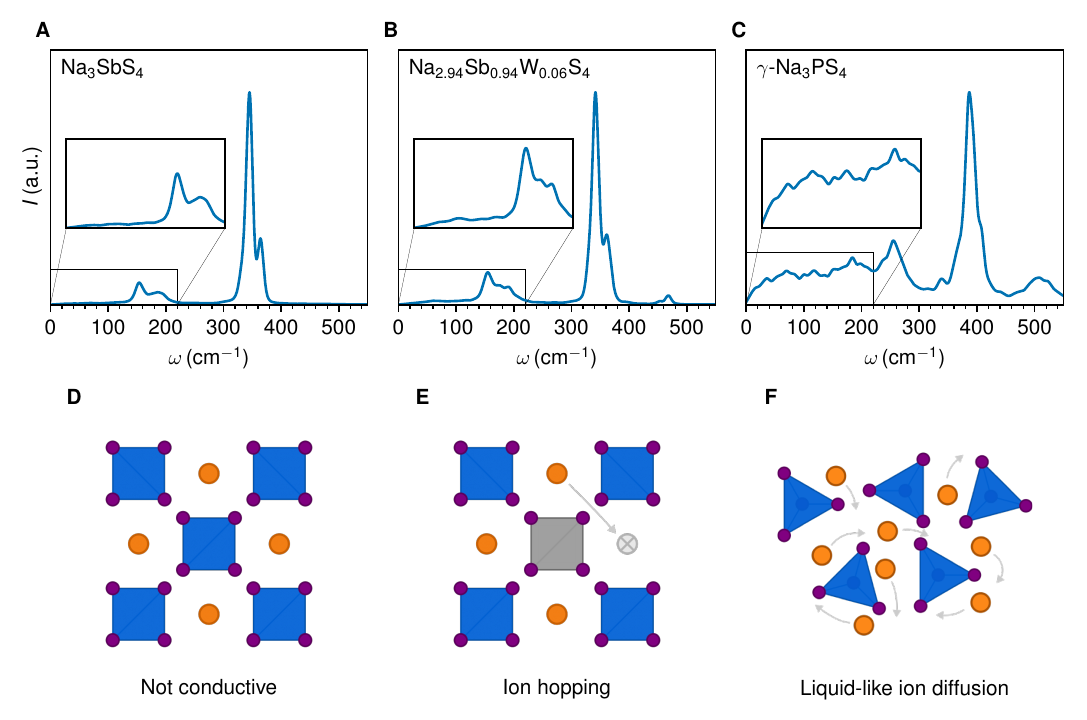}
    \caption{
      (\textbf{A}--\textbf{C}) Raman spectra of pristine and W-doped Na$_3$SbS$_4$
      at 300\,K as well as $\gamma$-Na$_3$PS$_4$ at 900\,K. The low-frequency regions are magnified to highlight the differences.
      (\textbf{D}--\textbf{F}) Schematic sketches of the structures and conduction mechanisms of these materials.}
    \label{fig:Na3SbS4}
\end{figure*}

The hypothesis that fast ionic conductors exhibit intense low-frequency Raman features prompts several questions, for instance, whether this behavior is universal across diverse diffusive systems, or whether it depends on the magnitude of the diffusion constant and the underlying transport mechanism. 
Having established the accuracy of our ML-accelerated Raman approach, we now address these questions for sodium pnictochalcogenides, Na$_3$PnS$_4$ (Pn = P, Sb). 
This family of Na-ion conducting materials is ideally suited for such an analysis, because it exhibits a rich variety of ion migration pathways that can be tuned through their ionic composition~\cite{gupta_etal_2021, brenner_etal_2022, maus_etal_2023}. 

Specifically, in \ch{Na3SbS4} superionic conduction can be induced via aliovalent doping \cite{chu2016room, zhang2016vacancy, hayashi2019sodium, jalem_etal_2020}. 
For example, W-doped Na$_{3-x}$Sb$_{1-x}$W$_x$S$_4$ has been identified as a composition exhibiting both high Na$^+$ conductivity and excellent material stability \cite{fuchs2019defect, fuchs_etal_2020, feng2021heavily, klarbring2024vacancy}.
In this case, Na$^+$ ion conduction is activated through the formation of Na$^+$ vacancies. 
Such are not present by the same amount in the pristine compound and their absence renders the undoped material non-conductive~\cite{bo2016computational, moon2018vacancy, miyagawa_etal_2024}. 
The resulting hopping-based conduction mechanism active in Na$_{3-x}$Sb$_{1-x}$W$_x$S$_4$~\cite{fuchs_etal_2020, feng2021heavily, klarbring2024vacancy} is notably distinct from the liquid-like conduction observed in AgI (see Fig.~\ref{fig:1}B). 

Interestingly, at high enough temperature the twin compound sodium thiophosphate in its pristine form transforms into a plastic polymorph, $\gamma$-\ch{Na3PS4}.
MD simulations showed that the transformation into the $\gamma$-phase is accompanied by liquid-like Na$^+$ motion and a marked increase in ionic conductivity ~\cite{famprikis_etal_2019a}. 
Complementary Raman measurements revealed the emergence of a central peak characteristic of dynamic symmetry breaking that accompanies the onset of liquid-like Na$^+$ diffusion in the $\gamma$-phase~\cite{famprikis_etal_2021}. 
Applying our ML-accelerated Raman approach to these materials therefore provides a unique opportunity to address the research questions posed above, namely to investigate how different ion migration mechanisms manifest in the Raman spectrum in general, and in its low-frequency features in particular.

To this end, we first compute Raman spectra of the pristine Na$_3$SbS$_4$ and W-doped Na$_{3-x}$Sb$_{1-x}$W$_x$S$_4$ ($x=6\%$) in order to isolate Raman signatures associated with Na$^+$ ion hopping (see Fig.~\ref{fig:Na3SbS4}A and C).
Both compounds exhibit well-defined Raman peaks near 350\,cm$^{-1}$, originating from vibrations of the SbS$_4$ units (see Fig.~\ref{fig:sup_vdos}). 
In the doped system, an additional peak appears at 470\,cm$^{-1}$, corresponding to WS$_4$ vibrations (see Fig.~\ref{fig:sup_vdos}). 
At first glance, the differences between the spectra are small, indicating that most vibrational modes, particularly those of the host lattice tetrahedra, are weakly affected by doping. 
Indeed, only subtle variations are observed in the low-frequency region, which appears to be mildly influenced by the hopping-based Na$^+$ conduction in the W-doped compound (see Fig.~\ref{fig:sup_MSD}). 

The striking absence of pronounced changes to Raman activity at low frequencies, despite the increased Na$^+$ diffusion (\textit{cf.}  Fig.~\ref{fig:Na3SbS4}D and E), can be attributed to the infrequent nature of ion hopping events. 
Specifically, hopping occurs roughly once every 5\,ps \cite{miyagawa_etal_2024}, which is rare compared with other dynamic events, such as Na$^+$ vibrations around their local equilibrium positions. 
Moreover, Na$^+$ hopping proceeds along well-defined ion migration paths and therefore does not break the crystalline symmetry to the extent of relaxing all Raman selection rules as is the case in highly disordered, liquid-like ion migration.
Taken together, these factors explain the modest changes observed in the low-frequency region of the Raman spectrum induced by Na$^+$ hopping. 

By contrast, our computed Raman spectrum for $\gamma$-\ch{Na3PS4} reveals pronounced and broad Raman signatures in the low-frequency regime, with substantial activity extending towards zero frequency.
Analysis of the VDOS (see Fig.~\ref{fig:sup_vdos}) shows that this low-frequency region is dominated by contributions from Na and even S atoms.
These result signify liquid-like motions of Na$^+$ ions (see Fig.~\ref{fig:Na3SbS4}F) and anharmonic vibrations of the host lattice arising from PS$_4$ tetrahedral rotations, consistent with previous studies discussed above~\cite{famprikis_etal_2019, famprikis_etal_2021}.
Taken together, the ion-conduction mechanism in $\gamma$-\ch{Na3PS4} shares key characteristics with that of AgI, namely liquid-like cation motion and relaxational host-lattice dynamics. 
Both break the crystalline symmetry on timescales relevant to Raman scattering and therefore relax the Raman selection rules, giving rise to pronounced low-frequency Raman intensity.

With this, we have addressed the research question posed above concerning possible interconnections between pronounced low-frequency Raman intensity and ionic conductivity.
While any diffusive system exhibits non-zero VDOS at zero frequency, a relaxation of Raman selection rules is required such that this translates into notable Raman intensity at zero frequency.
Our Raman predictions across the three sodium pnictochalcogenides -- one being non-conductive, one exhibiting hopping-based conduction, and one liquid-like conduction -- demonstrate that the lifting of Raman selection rules occurs only when cation motion is liquid-like and the host-lattice dynamics are relaxational.
Pronounced low-frequency Raman intensity thus emerges as a powerful indicator for fast ionic conduction, appearing exclusively in materials exhibiting liquid-like and therefore exceptionally rapid ionic transport. 
The ML-accelerated workflow for Raman predictions developed in this work is therefore particularly valuable for high-throughput materials screening, where it can assist in the interpretation of experimental data and enable standalone exploration of new material candidates.

Finally, we place our findings in the broader context of other materials in which Raman central peaks have been reported.
This comparison is important because Raman central peaks and related low-frequency features can arise from a broad spectrum of microscopic mechanisms that dynamically break crystalline symmetry and relax Raman selection rules on the timescale relevant to Raman scattering.
For instance, in classic oxide ferroelectrics such as KTa$_{1-x}$Nb$_x$O$_3$ and K$_{1-x}$Li$_x$TaO$_3$ they appear near to their second-order phase transitions~\cite{vugmeister1999second}. 
By contrast, in lead-halide perovskites the Raman central peak persists across a wide temperature range and is attributed to polarizability fluctuations that accompany disordered, localized and overdamped octahedral vibrations~\cite{yaffe2017local,menahem2023disorder,wu_etal_2017,gehrmann_egger_2019,zhu_etal_2022,caicedo-davila_etal_2024,ferreira_etal_2020,stock_etal_2020,weadock_etal_2020,lanigan-atkins_etal_2021,weadock_etal_2023,fransson_etal_2023,zhu_egger_2025}.
In superionic conductors such as the investigated AgI, a central peak has long been recognized as a characteristic feature of the high-temperature superionic phase. 
In an early model system, Geisel treated Ag$^+$ ion motion as Brownian diffusion in a sinusoidal potential and successfully reproduced the central peak via local polarizability fluctuations~\cite{geisel1977liquid}. 
Later, Brenner \textit{et al.} showed that the central peak in AgI originates from an overdamped tetrahedral oscillator mode coupled to rapid Ag$^+$ ion motions \cite{brenner_etal_2020}. 
Our results generalize this understanding to broader classes of solid electrolytes, where we identify a distinct low-frequency Raman response arising from the dynamic symmetry breaking and relaxation of Raman selection rules that accompany liquid-like cation motion.  
By establishing this link, our ML-accelerated Raman methodology not only rationalizes previous observations of central peaks in superionic phases such as AgI but also generalizes them into a predictive framework for identifying novel fast-ion conductors.


In summary, we have demonstrated that pronounced low-frequency Raman intensity serves as a robust spectroscopic signature of liquid-like ionic conduction in solid electrolytes.
Developing an ML-accelerated workflow that achieves near-ab initio accuracy at greatly reduced computational cost, we established a direct connection between Raman activity, ionic diffusivity, and the relaxation of selection rules caused by dynamic symmetry breaking.
Across representative sodium pnictochalcogenides, our results reveal that only systems exhibiting liquid-like cation motion and relaxational host-lattice dynamics display intense low-frequency Raman features, whereas systems exhibiting hopping-based ionic conduction do not.
By generalizing this understanding beyond the canonical case of AgI, we show that the breakdown of Raman selection rules provides a unifying framework for interpreting Raman central peaks across diverse material classes.
The ML-accelerated Raman approach introduced here thus not only rationalizes experimental observations of diffusive Raman scattering in superionic phases but also offers a predictive, data-efficient pathway for discovering and characterizing new fast-ion conductors.

\subsection*{Acknowledgments}

We thank Tomáš Bučko (Comenius University Bratislava) for valuable assistance in the initial phase of the project, and Tom Brenner, Matan Menahem and Omer Yaffe (Weizmann Institute of Science) for fruitful discussions and providing the experimental Raman spectrum of AgI.
Funding provided by Germany's Excellence Strategy (EXC 2089/1-390776260), by TUM.solar in the context of the Bavarian Collaborative Research Project Solar Technologies Go Hybrid (SolTech), and by the TUM-Oerlikon Advanced Manufacturing Institute are gratefully acknowledged.
The authors further acknowledge the Gauss Centre for Supercomputing e.V. for funding this project by providing computing time through the John von Neumann Institute for Computing on the GCS Supercomputer JUWELS at Jülich Supercomputing Centre.

\bibliographystyle{apsrev4-2}
\bibliography{references}

\subsection*{Data availability}

The raw data for all calculations are available on Zenodo (\url{http://doi.org/10.5281/zenodo.17703839}).

\clearpage

\renewcommand{\thefigure}{S\arabic{figure}}
\renewcommand{\thetable}{S\arabic{table}}
\renewcommand{\theequation}{S\arabic{equation}}
\renewcommand{\thepage}{S\arabic{page}}
\setcounter{figure}{0}
\setcounter{table}{0}
\setcounter{equation}{0}
\setcounter{page}{1}

\title{Supplemental Material: Revealing Fast Ionic Conduction in Solid Electrolytes through Machine Learning
Accelerated Raman Calculations}
\maketitle

\subsection*{Materials and Methods}

We computed atomic trajectories from MD simulations with machine learning force fields (MLFFs) \cite{jinnouchi_etal_2019, jinnouchi_etal_2019a, jinnouchi_etal_2020} that were pre-trained on-the-fly based on DFT calculations in VASP \cite{ kresse_furthmuller_1996a}.
For AgI,  Na$_3$SbS$_4$, and Na$_{3-x}$Sb$_{1-x}$W$_x$S$_4$, we extended existing trajectories from our previous study \cite{miyagawa_etal_2024}.
For $\gamma$-Na$_3$PS$_4$, we created a new MLFF and generated an additional MLMD trajectory.
The underlying DFT calculations were performed with the PBE functional and PAW pseudopotentials.
For Na$^+$, a potential corresponding to a valence configuration with seven electrons (2p$^6$, 3s$^1$) was used.
The plane wave energy cutoff was set to 280\,eV for AgI and 400\,eV for the Na-based materials.

For AgI, we performed an MD simulation with a supercell containing 128 atoms at a temperature of 500\,K. A timestep of 10\,fs was used. After an initial equilibration phase of 10\,ps, the production trajectory covered 120\,ps. 
This length includes our extension of the original simulation.
For Na$_3$SbS$_4$, we used a supercell with 128 atoms at a temperature of 300\,K. 
The timestep was set to 2\,fs. 
The length of the equilibration phase was 5\,ps, and the length of the extended production run was 100\,ps.
The same settings were also used for the doped Na$_{3-x}$Sb$_{1-x}$W$_x$S$_4$. 
However, in this case the cell only contained 127 atoms due to the presence of the Na$^+$ vacancy.
For $\gamma$-Na$_3$PS$_4$, we used a larger supercell with 256 atoms, at a temperature of 900\,K. 
The timestep was 2\,fs, the equilibration phase was 5\,ps, and the length of the production run was 100\,ps.
Temperature was controlled using the Nosé-Hoover thermostat in all cases~\cite{nose1984unified, hoover1985canonical}.

Raman spectra were computed from polarizability timeseries containing polarizability tensors for each 10th configuration from the MD trajectories \cite{thomas_etal_2013}.
Specifically, the spherically-averaged total Raman intensity over all polarization directions was computed, involving the expression $45a_\tau^2 + 7\gamma_\tau^2$ where the $a_\tau$ and $\gamma_\tau$ denote the frequency-dependent tensor invariants obtained via autocorrelation.
The full expression for computing the spectra is given in eq.~\ref{eq:md_raman}.
The resulting spectra were smoothened in order to reduce noise. 
This was done via a convolution with a Gaussian smoothing function with $\sigma=4\,\mathrm{cm}^{-1}$.

The polarizability tensors were obtained using an equivariant ML approach based on $\lambda$-SOAP \cite{grisafi_etal_2018}.
This was implemented using the featomic and metatensor packages \cite{metatensor}.
Learning curves were obtained by training the model several times with training data sets of different sizes and evaluating its performance on a separate test set.
The training and test data sets consisted of reference polarizabilities for randomly selected snapshots from the MD trajectories.
The size of the final training data set used to train the production model for obtaining the Raman spectrum was $N_\mathrm{T} = 300$ for each material.
The convergence of the spectra with respect to training set size is shown in
Figs.~\ref{fig:sup_training_convergence_AgI}--\ref{fig:sup_training_convergence_gNa3PS4},
and the convergence with respect to trajectory length is shown in
Figs.~\ref{fig:sup_length_convergence_AgI}--\ref{fig:sup_length_convergence_gNa3PS4}.

The reference polarizability values required for training and performance evaluation were obtained from first-principles dielectric tensor calculations using density functional perturbation theory \cite{baroni_resta_1986}, as implemented in VASP \cite{gajdos_etal_2006}.
These calculations were performed with the same PAW potentials as the corresponding MD simulations.
Note that the timeseries used for the Raman calculations consist entirely of ML predictions, which means they do not directly contain any of the DFPT reference polarizability tensors.

\subsection*{Expressions for Raman Calculations}

The full expression we used for computing the Raman spectrum is
\begin{equation}
  \label{eq:md_raman}
  I(\omega) \propto
  \frac{(\omega_\text{in} - \omega)^4}{\omega} \,
  \frac{45a_\tau^2 + 7\gamma_\tau^2}{45}.
\end{equation}
Here, $a_\tau$ denotes the trace of the autocorrelated timeseries
\begin{equation}
\begin{aligned}
  a^2_{\tau}
  = \frac{1}{9}
  \int_{-\infty}^{\infty}
  & \mathrm{d} t \, e^{-i\omega t} \,\,
  \Bigl \langle
     \bigl( \dot{\alpha}_{xx}(\tau) + \dot{\alpha}_{yy}(\tau)
      + \dot{\alpha}_{zz}(\tau)\bigr) \\
    & \bigr(\dot{\alpha}_{xx}(\tau+t) + \dot{\alpha}_{yy}(\tau+t)
      + \dot{\alpha}_{zz}(\tau+t)\bigr)
  \Bigr \rangle_{\tau}
\end{aligned}
\label{eq:a_tau}
\end{equation}
and $\gamma_\tau$ denotes the anisotropy
\begin{widetext}
\begin{equation}
\begin{aligned}
  \gamma^2_{\tau}
  &= 3\int \mathrm{d} t \, e^{-i \omega t} \,\,
  \Bigl\langle
      \dot{\alpha}_{xy}(\tau)\, \dot{\alpha}_{xy}(\tau+t)
    + \dot{\alpha}_{yz}(\tau)\, \dot{\alpha}_{yz}(\tau+t)
    + \dot{\alpha}_{zx}(\tau)\, \dot{\alpha}_{zx}(\tau+t)
  \Bigr\rangle_{\tau} \\
  & \qquad + \frac{1}{2}\int \mathrm{d} t \, e^{-i\omega t} \,\,
  \Bigl\langle
    \left(\dot{\alpha}_{xx}(\tau)- \dot{\alpha}_{yy}(\tau)\right ) \,
    \left(\dot{\alpha}_{xx}(\tau+t) - \dot{\alpha}_{yy}(\tau+t)\right)
  \Bigr\rangle_{\tau} \\
  & \qquad + \frac{1}{2}\int \mathrm{d} t \, e^{-i\omega t} \,\,
  \Bigl\langle
    \left ( \dot{\alpha}_{yy}(\tau)- \dot{\alpha}_{zz}(\tau)\right ) \,
    \left(\dot{\alpha}_{yy}(\tau+t) - \dot{\alpha}_{zz}(\tau+t)\right)
  \Bigr\rangle_{\tau} \\
  & \qquad + \frac{1}{2}\int \mathrm{d} t \, e^{-i\omega t} \,\,
  \Bigl\langle
    \left ( \dot{\alpha}_{zz}(\tau)- \dot{\alpha}_{xx}(\tau)\right ) \,
    \left(\dot{\alpha}_{zz}(\tau+t) - \dot{\alpha}_{xx}(\tau+t)\right)
  \Bigr\rangle_{\tau}.
\end{aligned}
\end{equation}
\end{widetext}
Note that we did non include the Bose--Einstein weighting factor in the computation of the spectra.

\subsection*{Supplementary Figures}

\begin{figure*}
    \centering
    \includegraphics[width=0.8\textwidth]{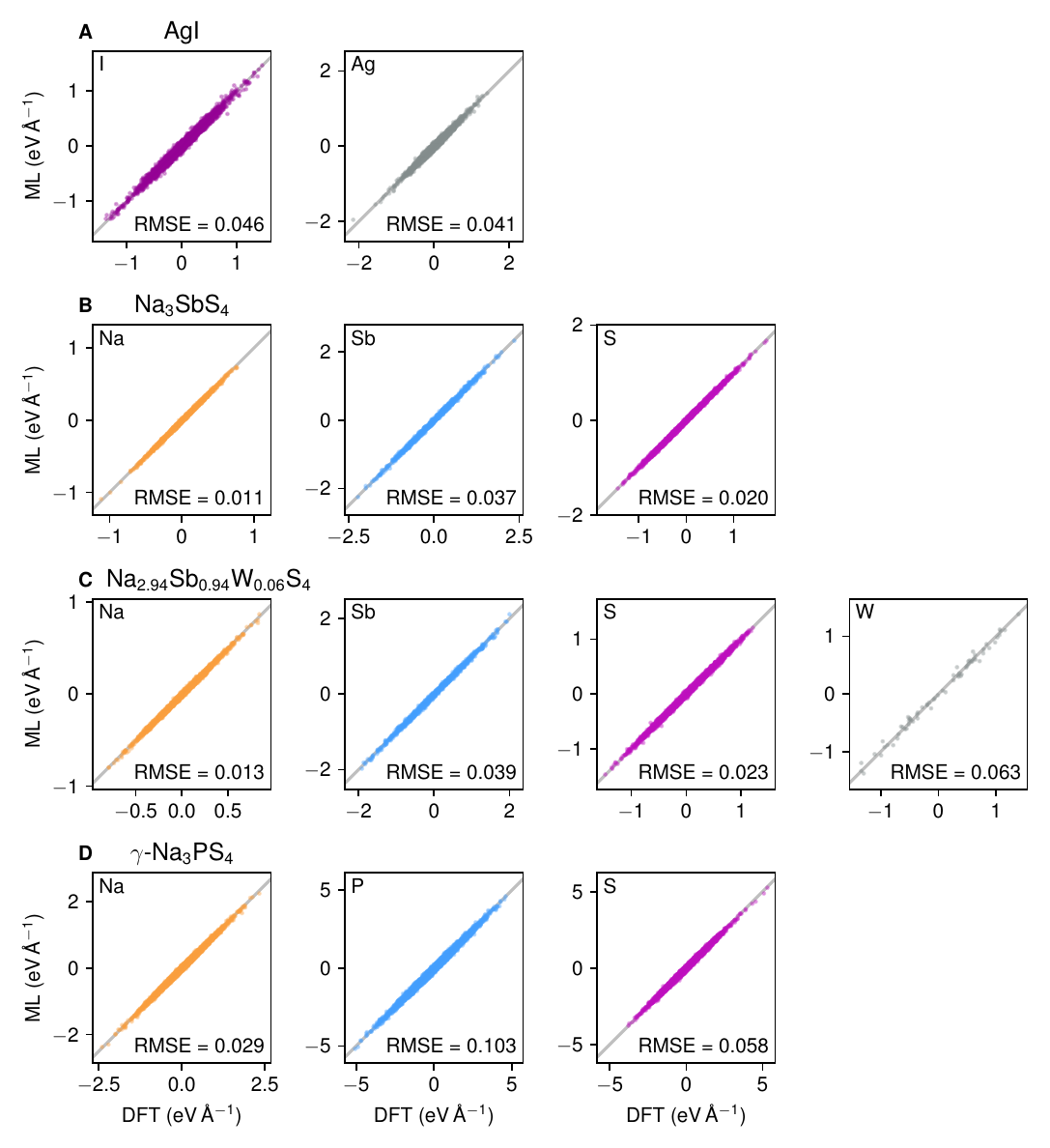}
    \caption{\textbf{Forces predicted by the ML force fields compared to DFT reference values.}
        (\textbf{A}) AgI.
        (\textbf{B}) Na$_3$SbS$_4$.
        (\textbf{C}) Na$_{2.94}$Sb$_{0.94}$W$_{0.06}$S$_4$.
        (\textbf{D}) $\gamma$-Na$_3$PS$_4$. Root mean squared errors (RMSE) are given in units of eV\,\AA$^{-1}$.
    }
    \label{fig:sup_mlff_scatterplots}
\end{figure*}

The prediction performance of the MLFFs is shown in Fig.~\ref{fig:sup_mlff_scatterplots} by comparing MLFF predictions for the forces to DFT reference values on a per-atom basis. 
The root mean square error (RMSE) is less than 0.1\,eV\,\AA$^{-1}$ in all cases except for the high-temperature simulation of $\gamma$-N$_3$PS$_4$.

\begin{figure*}
    \centering
    \includegraphics[width=0.8\textwidth]{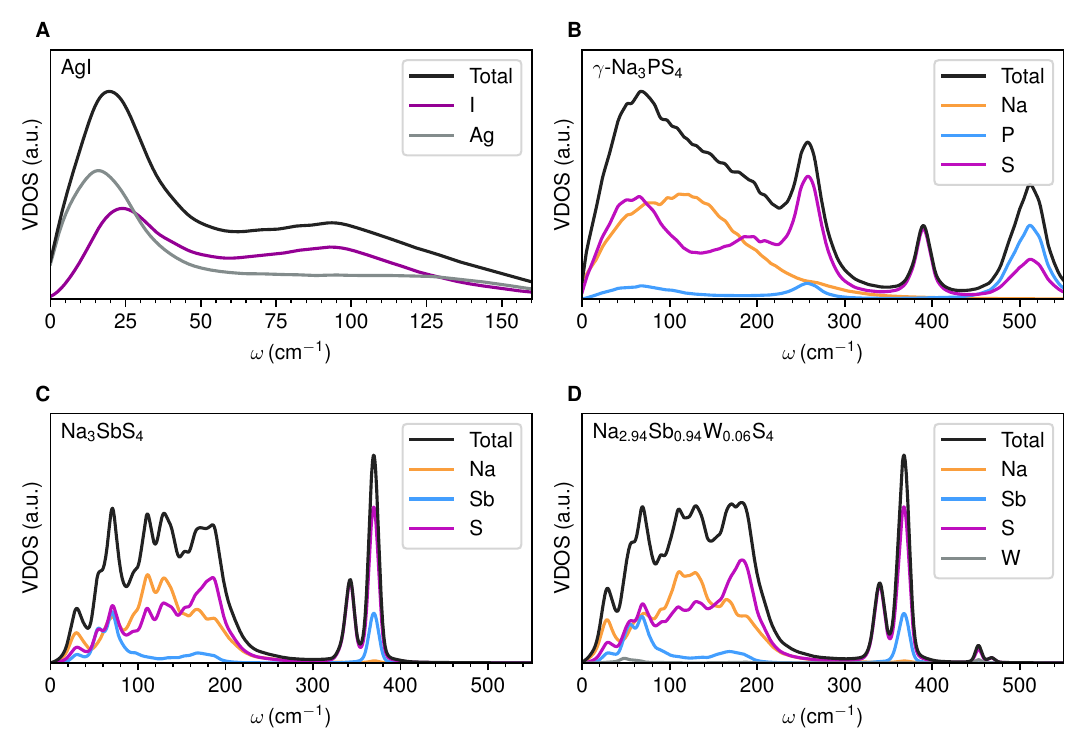}
    \caption{\textbf{Atom-resolved vibrational density of states (VDOS).}
        (\textbf{A}) AgI.
        (\textbf{B}) Na$_3$SbS$_4$.
        (\textbf{C}) Na$_{2.94}$Sb$_{0.94}$W$_{0.06}$S$_4$.
        (\textbf{D}) $\gamma$-Na$_3$PS$_4$.
    }
    \label{fig:sup_vdos}
\end{figure*}

The atom-resolved vibrational density of states obtained from the MD simulations is shown for all investigated systems in Fig.~\ref{fig:sup_vdos}. 
This allows to distinguish which atoms contribute to vibrations at which frequencies. 
The mobile ions primarily contribute to the low-frequency region in all cases. 
Also note that in the case of $\gamma$-Na$_3$PS$_4$, both the Na and the S atoms contribute to the zero-frequency limit.
This is related to ionic diffusion and to the rotational motion of the PS$_4$ tetrahedra.

\begin{figure*}
    \centering
    \includegraphics[width=0.8\textwidth]{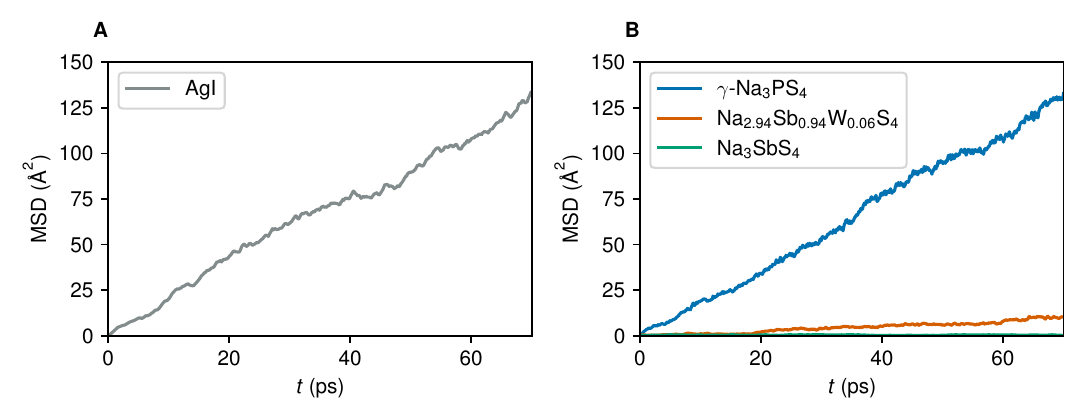}
    \caption{\textbf{Mean square displacements of the mobile ions.}
        (\textbf{A}) AgI.
        (\textbf{B}) Na-based materials.}
    \label{fig:sup_MSD}
\end{figure*}

Mean square displacements (MSD) of the mobile ions are shown in Fig.~\ref{fig:sup_MSD}. 
This demonstrate that disordered materials $\gamma$-Na$_3$PS$_4$ shows significantly higher diffusivity than the hopping-based conductor Na$_{2.94}$P$_{0.94}$W$_{0.06}$S$_4$.
The MSD in $\gamma$-Na$_3$PS$_4$ reaches similar values to AgI, despite belonging to a different material class.

\begin{figure*}
    \centering
    \includegraphics{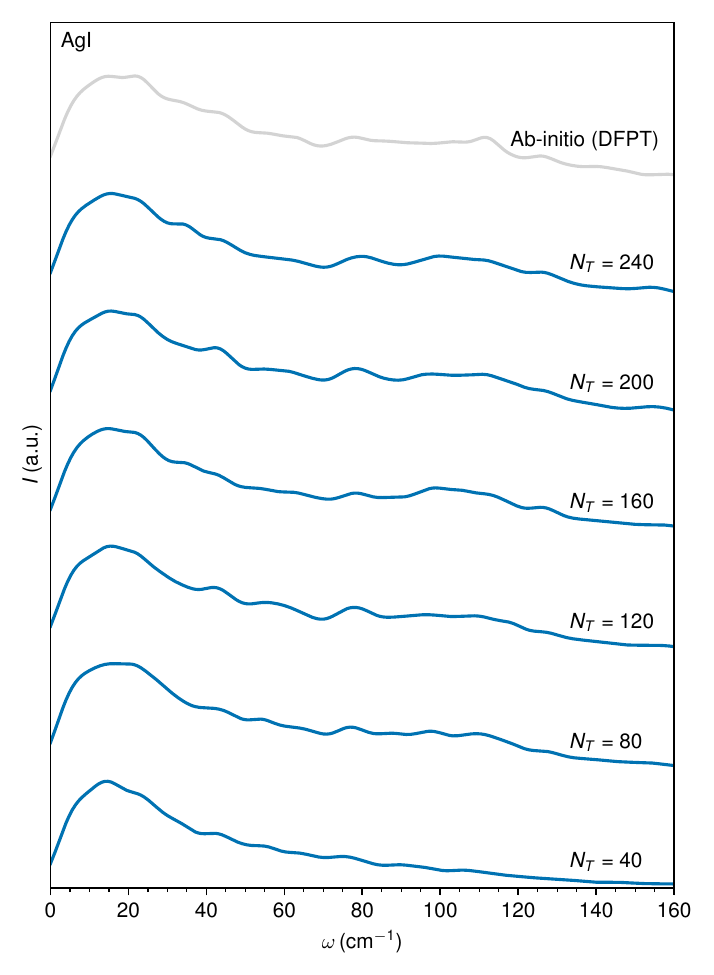}
    \caption{\textbf{Computed Raman spectra for AgI with different training set sizes.}}
    \label{fig:sup_training_convergence_AgI}
\end{figure*}

\begin{figure*}
    \centering
    \includegraphics{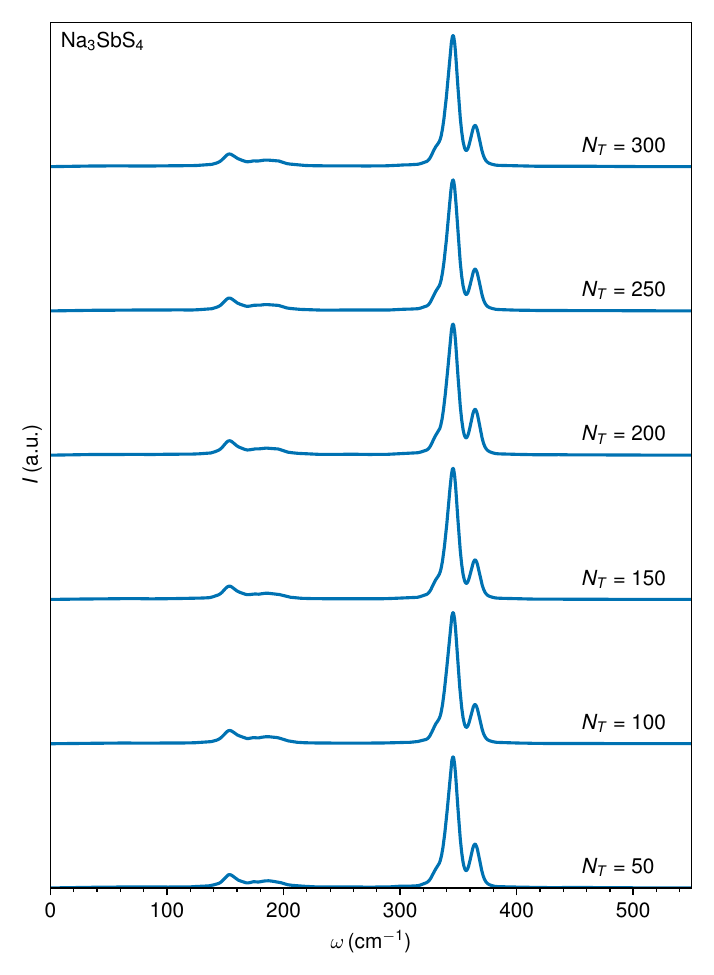}
    \caption{\textbf{Computed Raman spectra for Na$_3$SbS$_4$ with different training set sizes.}}
    \label{fig:sup_training_convergence_Na3SbS4}
\end{figure*}

\begin{figure*}
    \centering
    \includegraphics{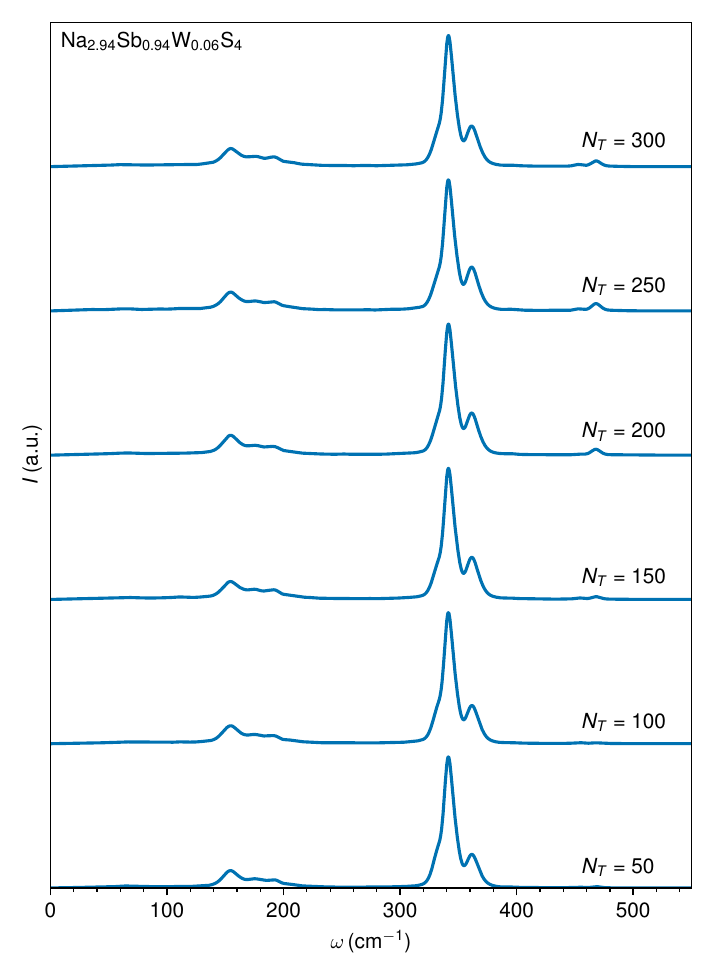}
    \caption{\textbf{Computed Raman spectra for Na$_{2.94}$Sb$_{0.94}$W$_{0.06}$S$_4$ with different training set sizes.}}
    \label{fig:sup_training_convergence_Na3SbS4+W}
\end{figure*}

\begin{figure*}
    \centering
    \includegraphics{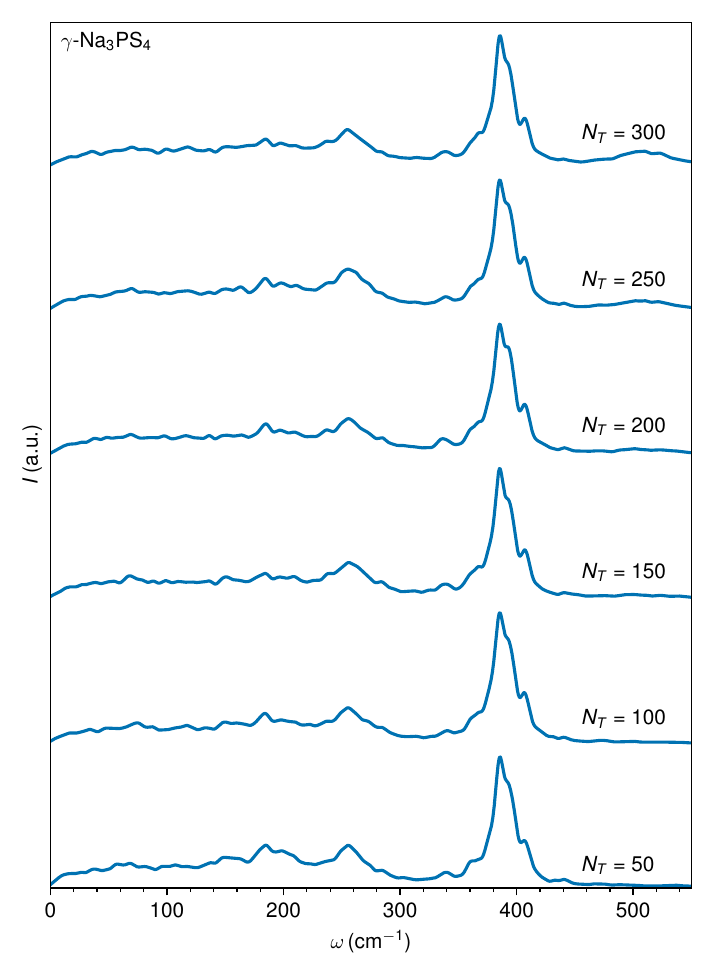}
    \caption{\textbf{Computed Raman spectra for $\gamma$-Na$_3$PS$_4$ with different training set sizes.}}
    \label{fig:sup_training_convergence_gNa3PS4}
\end{figure*}

\begin{figure*}
    \centering
    \includegraphics{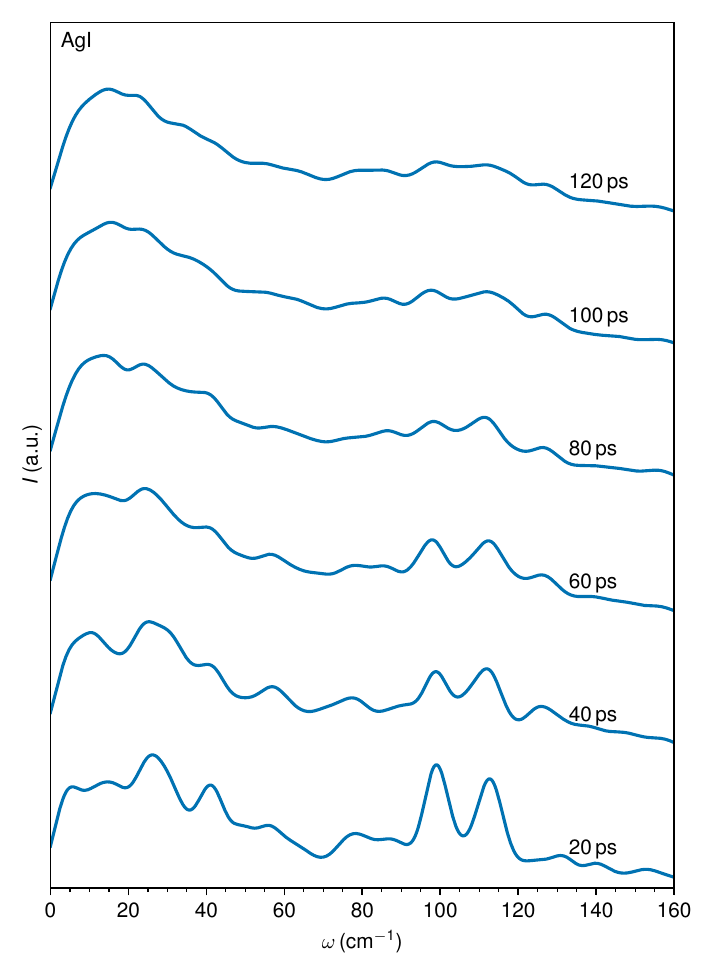}
    \caption{\textbf{Computed Raman spectra for AgI using subtrajectories of different lengths.}}
    \label{fig:sup_length_convergence_AgI}
\end{figure*}

\begin{figure*}
    \centering
    \includegraphics{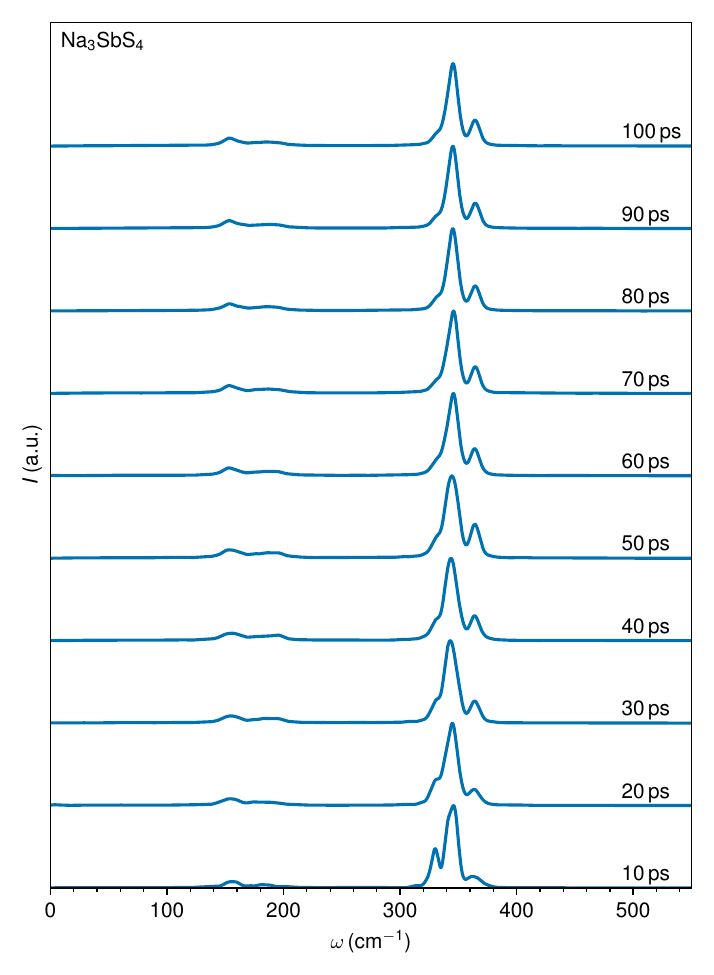}
    \caption{\textbf{Computed Raman spectra for Na$_3$SbS$_4$ using subtrajectories of different lengths.}}
    \label{fig:sup_length_convergence_Na3SbS4}
\end{figure*}

\begin{figure*}
    \centering
    \includegraphics{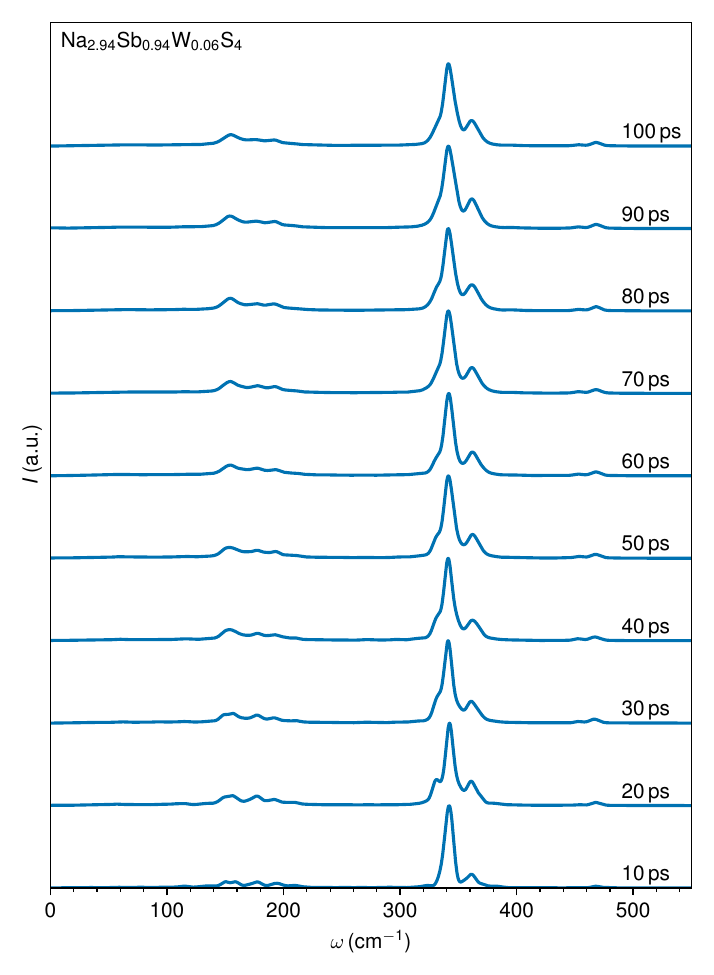}
    \caption{\textbf{Computed Raman spectra for Na$_{2.94}$Sb$_{0.94}$W$_{0.06}$S$_4$ using subtrajectories of different lengths.}}
    \label{fig:sup_length_convergence_Na3SbS4+W}
\end{figure*}

\begin{figure*}
    \centering
    \includegraphics{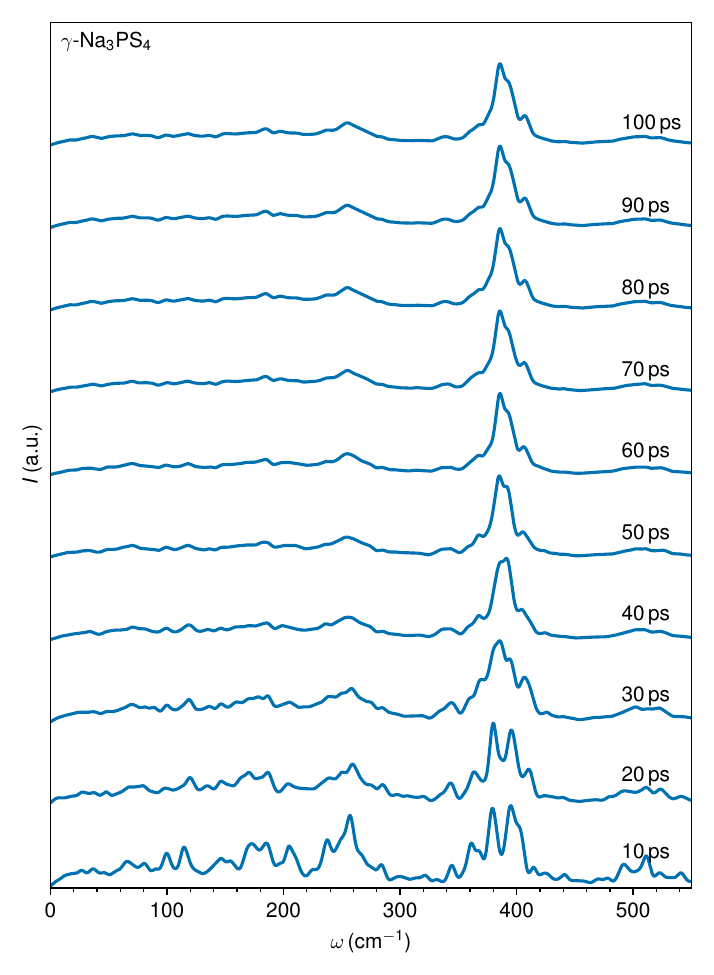}
    \caption{\textbf{Computed Raman spectra for $\gamma$-Na$_3$PS$_4$ using subtrajectories of different lengths.}}
    \label{fig:sup_length_convergence_gNa3PS4}
\end{figure*}

The convergence of the spectra with respect to the size of the training data set and the total length of the trajectory is an important aspect to consider.
Figs.~\ref{fig:sup_training_convergence_AgI}--\ref{fig:sup_training_convergence_gNa3PS4} contain spectra obtained with different training set sizes.
This comparison demonstrates that the sizes we chose are sufficient to obtain converged spectra.
Figs.~\ref{fig:sup_length_convergence_AgI}--\ref{fig:sup_length_convergence_gNa3PS4} contain spectra obtained with different subtrajectory lengths.
These spectra were obtained with the same training set sizes as the results shown in the main text.
Note that using a long trajectory is particularly important in the case of Na$_{2.94}$Sb$_{0.94}$W$_{0.06}$S$_4$.
This is due to the fact that ion hopping in this material is a relatively rare event, such that the long trajectory is required in order to sample a sufficient number of hopping events.
All spectra shown in these figures were smoothed using Gaussian smoothing with $\sigma = 4\,\mathrm{cm}^{-1}$.

\begin{figure*}
    \centering
    \includegraphics[width=0.8\textwidth]{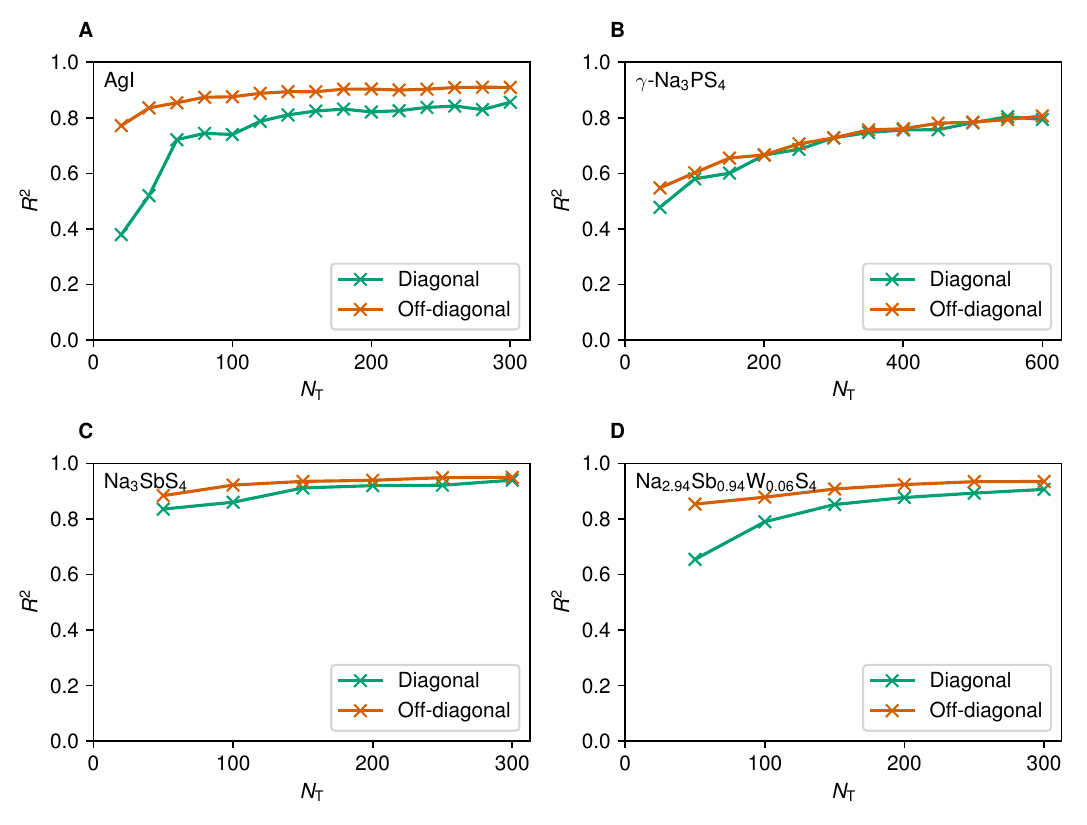}
    \caption{\textbf{Learning curves of the polarizability ML models.}
        (\textbf{A}) AgI.
        (\textbf{B}) Na$_3$SbS$_4$.
        (\textbf{C}) Na$_{2.94}$Sb$_{0.94}$W$_{0.06}$S$_4$.
        (\textbf{D}) $\gamma$-Na$_3$PS$_4$.
    }
    \label{fig:sup_learning}
\end{figure*}

\end{document}